# Energy Momentum Localization in Quantum Gravity

Stuart Marongwe [1*]

[1] University of Botswana
* Correspondence: stuartmarongwe@gmail.com

**Abstract:** We introduce quantum spatio-temporal dynamics (QSD) as modeled by the Nexus Paradigm (NP) of quantum gravity to resolve the problem of energy- momentum localization in a gravitational field. Currently, the gravitational field as described using the language of geometry modeled under General Relativity (GR) fails to provide a generally accepted definition of energy-momentum. Attempts at resolving this problem using geometric methods have resulted in various energy-momentum complexes whose physical meaning remain dubious since the resulting complexes are non-tensorial under a general coordinate transformation. In QSD, the tangential manifold is the affine connection field in which energy-momentum localization is readily defined. We also discover that the positive mass condition is a natural consequence of quantization and that dark energy is a Higgs like field with negative energy density everywhere. Finally, energy-momentum localization in quantum gravity shows that a free falling object will experience larger vacuum fluctuations (uncertainties in location) in strong gravity than in weak gravity and that the amplitudes of these oscillations define the energy of the free falling object.

## 1. Introduction

Since the very inception of GR, Einstein was aware of the energy inherent in a gravitational field and that this energy must also gravitate [1-2]. Thus in GR gravity must self-gravitate. Self-gravitation of the gravitational field is also a problem in perturbative approaches to quantum gravity as it leads to infinities in the strong gravity regime [3-5]. Einstein failed to find a symmetric tensor that would properly localize the energy-momentum of the gravitational field but instead introduced a non-covariant pseudo-tensor. Pseudo-tensors are non-tensorial under a coordinate transformation and therefore the problem of energy localization remains unresolved. Solving energy-momentum localization in gravity will provide answers to the enigmatic sources of dark energy (DE) and dark matter (DM) as argued by Nash in Ref.[1]. A quantum gravity (QG) solution would provide deeper understanding of these enigmas from a Quantum Field Theory (QFT) perspective.

The search for a solution to this problem has been an area of active research. A brief survey of the literature shows notable attempts using super-energy tensors [6-8], quasi local expressions [9-12] and energy momentum complexes of Einstein [13-14],

Papapetrou [15], Møller [16], Bergman-Thompson [17] Landau- Lifshitz [18] and Weinberg[19]. As highlighted by Randinschi *et al* [20] pseudo-tensors have an inherent central problem which is their coordinate dependence. Further, their construction involves two parts one for matter and one for the gravitational field and herein lies the problem which is embedded in their very mathematical construction. From a physical perspective, the equivalence principle makes no distinction between gravitational mass and inertial mass. The gravitational mass is associated with the gravitational field and so it is imperative to find an expression in which the energy of the gravitational field is expressed in terms of the gravitational mass embedded within it and vice versa. That is, we seek an expression in which the stress-momentum tensor in Einstein's field equation is expressed in terms of the field momentum.

The paper is structured as follows: first we introduce a semi-classical derivation of quantum gravity and the guiding principles for the quantization process. We then introduce Hamilton's Ricci flow which enables the derivation of a covariant canonical formulation of quantum gravity. This formulation then enables energy-momentum localization in quantum gravity. The final section is a discussion of the energy-momentum localization.

**2.0 A semi classical derivation of quantum gravity.**

The search for quantum gravity begins with adopting an intrinsic geometry in which Einstein's field equations are interpreted as describing curved world lines in flat space-time. By adopting this interpretation, one can start embarking on an alternative path to quantum gravity since QFT is a theory built on flat space-time and has curved lines that appear as a sum over histories in the Feynman interpretation of QFT. Moreover the Ricci curvature tensor in GR is the average of the possible paths a test particle can take in a gravitational field. That is, given two vector fields $X$ and $Z$, $Ric(X,Z) = \sum_i^n \langle R(X,e_i)Z, e_i \rangle = -\frac{1}{2}\Delta g_{ij}$. This statistical view which is analogous to thermal diffusion provides an intuitive glimpse in which Hamilton's Ricci Flow $\partial_t g_{\mu\nu} = \Delta g_{\mu\nu}$ [21] plays an important role in the formulation of quantum gravity. In the subsequent section, we introduce the Ricci Flow and its basic properties which will play a pivotal role in the formulation of a self-consistent model of QG.

**2.1 Hamilton's Ricci Flow**

In 1982 Richard Hamilton introduced the concept of Ricci flow [22]. In a Riemannian manifold $(M, g)$, the Ricci Flow is partial differential equation that evolves the metric tensor

$$\partial_t g_{\mu\nu} = -2Ric\left(g_{\mu\nu}(t)\right). \tag{1}$$

where $Ric(g_{\mu\nu}(t))$ denotes the Ricci curvature of the metric $g_{\mu\nu}(t)$. The time evolution of the metric under the Ricci flow spreads the curvature evenly through space. It should be noted that the Ricci Flow also includes a quadratic reaction term which will be included later in the work.

In compact Einstein manifolds the Ricci Flow is expressed as follows:

$$\partial_t g_{\mu\nu} = -\frac{cr_H}{3} G_{\mu\nu}. \tag{2}$$

where $G_{\mu\nu} = R_{\mu\nu} - \frac{1}{2}Rg_{\mu\nu}$ is the Einstein tensor, $c$ the speed of light and $r_H$ the Hubble radius. Compact Einstein manifolds have the form

$$G_{\mu\nu} = kg_{\mu\nu}. \tag{3}$$

In GR with the cosmological constant $\Lambda$, the compact Einstein manifold assumes the form

$$G_{\mu\nu} = \Lambda g_{\mu\nu}. \tag{4}$$

Therefore, vacuum solutions of Einstein's field equations are compact Einstein manifolds with $k$ proportional to the cosmological constant. The above equation describes a Ricci soliton of De Sitter topology and is divergenceless. That is $\nabla_\mu G_{\mu\nu} = \Lambda \nabla_\mu g_{\mu\nu} = \kappa \rho_{DE} \nabla_\mu g_{\mu\nu} = 0$ where $\kappa$, is the Einstein constant and $\rho_{DE}$ is the dark energy density. Here we have a packet of localized vacuum energy in the form of a Ricci soliton in which the energy conservation holds. In other words, the Ricci soliton is a self-gravitating gravitational field.

## 2.2 Space-time Quantization

In the Nexus Paradigm of quantum gravity, we begin the quantization process by considering a large but finite patch of Minkowski space equipped with a non-degenerate symmetric bilinear form on the tangent space. We adopt a local coordinate system to avoid the need of an origin as well as point like events which are the sources of divergences in QFT. The local coordinate system makes the Minkowski space a displacement vector space. The inner product is therefore

$$\Delta s^2 = \Delta x^\mu \Delta x_\mu = \Delta x^2 + \Delta y^2 + \Delta z^2 - c^2 \Delta t^2$$

$$= (A\Delta x + B\Delta y + C\Delta z + icD\Delta t)(A\Delta x + B\Delta y + C\Delta z + icD\Delta t). \quad (5)$$

Upon multiplying the right side we note that to get all the cross terms such as $\Delta x \Delta y$ to vanish we must assume

$$AB + BA = 0, \cdots \text{ and } A^2 = B^2 = \cdots = 1. \quad (6)$$

The above conditions generate a Clifford Algebra which implies that the coefficients $(A, B, C, D)$ must be matrices, specifically the Dirac gamma matrices. These matrices are square roots of the Minkowski metric

$$\gamma^\mu \gamma^\nu = \eta^{\mu\nu}. \quad (7)$$

Thus the displacement vectors $\Delta x^\mu = a\gamma^\mu$ reside in Clifford space $Cl_{1,3}(R)_C$ implying an intrinsic quantized spin and can be perceived as Dirac 4-vector matrices analogous to the Pauli vector matrices. They are also quantized wave packets of space-time and can be expressed as Fourier integrals as follows

$$\Delta x_n^\mu = \frac{2r_{HS}}{n\pi} \gamma^\mu \int_{-\infty}^{\infty} sinc(kx) e^{ikx} dk$$

$$= \gamma^\mu \int_{-\infty}^{\infty} a_{nk} \varphi_{(nk)} dk. \quad (8)$$

where $\quad \frac{2r_{HS}}{n\pi} = \sum_{k=-\infty}^{k=+\infty} a_{nk}. \quad (9)$

Here $r_{HS}$ is the Hubble radius, $\varphi_{(nk)} = sinc(kx)e^{ikx}$ are Bloch energy eigenstate functions in which the four wave vectors assume the following quantized values

$$k^\mu = \frac{n\pi}{r_{HS}^\mu} \quad n = \pm 1, \pm 2 \ldots 10^{60}. \quad (10)$$

We set a high energy cut off limit at the Planck 4-length since no measurement can be obtained below this length without the creation of a black hole and the low energy cut off limit being the Hubble 4-radius since no information can be obtained beyond the cosmic horizon. The $10^{60}$ states arise from the ratio of Hubble four radius to the Planck four length. The Bloch functions in each eigenstate of space-time generate an infinite Bravais four lattice. The conjugate momentum for the displacement vectors is

$$\Delta p_n^\mu = \frac{2np_1}{\pi} \gamma^\mu \int_{-\infty}^{\infty} \varphi(kx) dx$$

$$= \gamma^\mu \int_{-\infty}^{\infty} c_n \varphi(kx) dx. \quad (11)$$

Where $p_1$ is the four momentum of the ground state

The wave packet is essentially a particle of four-space and can be envisioned as enveloping a spherically symmetric lump of energy from the quantum vacuum. This vacuum energy can be in any form of the fields described by the Standard Model of particle physics.

We seek to find the relationship between these wave packets of spacetime and the Ricci solitons of Eq.(4). First we determine the norm squared of the four momentum of the $n$-th state wave packet. We compute this norm by multiplying the inner product of Eq.(10) by the square of the reduced Planck constant

$$(\hbar)^2 k^\mu k_\mu = \frac{E_n^2}{c^2} - \frac{3(nhH_0)^2}{c^2} = 0. \tag{12}$$

where $H_0$ is the Hubble constant. We then express Eq.(12) in terms of the cosmological constant, $\Lambda$ as

$$\Lambda_n = \frac{E_n^2}{(hc)^2} = \frac{3k_n^2}{(2\pi)^2} = n^2 \Lambda. \tag{13}$$

From Eq.(13), the wave packet can be considered as a compact Einstein manifold or a trivial Ricci soliton of positive Ricci curvature expressed in the form

$$G_{(nk)\mu\nu} = n^2 \Lambda g_{(n,k)\mu\nu} = n^2 \kappa \rho_\Lambda g_{(n,k)\mu\nu}. \tag{14}$$

Clearly Eq.(14) depicts a self-gravitating Ricci soliton and as explained in Refs: [23-26] this is DM which is a localized packet of vacuum energy $n^2 \rho_\Lambda$ in the $n$-th quantum state. Thus DM is a Ricci soliton and should exhibit the following soliton characteristics

1. It is a localized lump of (vacuum) energy
2. It preserves its form while growing or diminishing in size
3. It preserves its speed and form after collision with another soliton

The lowest quantum state from Eq.(10) occurs when $n = 1$ suggesting that for Eq.(14) to become Ricci flat in this state, a Ricci soliton in the ground state must be removed from the right side yielding Einstein's vacuum field equations in the quantized spacetime as

$$G_{(nk)\mu\nu} = (n^2 - 1)\Lambda g_{(n,k)\mu\nu} = (n^2 - 1)k\rho_\Lambda g_{(n,k)\mu\nu}. \tag{15}$$

The above equation depicts a decay mechanism in which a high energy graviton emits a ground state graviton to assume a low energy quantum state. The force exerted by the emission process can be computed via the Uncertainty Principle. We consider the ground state graviton as having a temporal interval $\Delta t$ equal to the Hubble time and a spatial interval $\Delta x$ equal to the Hubble radius

Thus
$$\Delta t \Delta E = \Delta t \Delta x . F = \frac{cF}{H_0^2} \sim \frac{h}{2\pi}$$

Therefore
$$F \sim \frac{hH_0}{c^2} \cdot \frac{H_0 c}{2\pi} = m_G \cdot a. \tag{16}$$

This implies that the mass of the ground state graviton is $m_G = \frac{hH_0}{c^2}$ which in 3D space is $m_G = \frac{3hH_0}{c^2}$ and the graviton induced acceleration is $a = \frac{H_0 c}{2\pi}$. Thus a spherical volume of space containing a dark energy mass $M_\Lambda(r)$ within a radius $r$ will always generate a scale invariant outward acceleration of $\frac{GM_\Lambda(r)}{r^2} = -\frac{H_0 c}{2\pi}$. This acceleration was first empirically observed by Milgrom from data on galaxy rotation curves [27]. As noted by Milgrom, non-Newtonian dynamics begins to manifest at this critical acceleration. This critical acceleration therefore marks a transition from the classical to the quantum gravity regime.

If the graviton field is perturbed by the presence of baryonic matter then Eq. (15) becomes

$$G_{(nk)\mu\nu} = kT_{\mu\nu} + (n^2 - 1)\Lambda g_{(nk)\mu\nu}. \tag{17}$$

From Ref.[23] the solution to Eq.(14) is computed as

$$ds^2 = -\left(1 - \left(\frac{2}{n^2}\right)\right)c^2 dt^2 + \left(1 - \left(\frac{2}{n^2}\right)\right)^{-1} dr^2 + r^2(d\theta^2 + \sin^2\theta d\varphi^2). \tag{18}$$

The above metric equation describes curved worldlines in flat spacetime and has no singularities nor divergencies. At high energies which are characterized by microcosmic scale wavelengths of the graviton and high values of $n$, the worldline is rectilinear and the local coordinates are highly compact or localized. This aspect also reveals asymptotic freedom in quantum gravity since for high values of $n$, gravity (world line curvature) vanishes asymptotically. Thus at high energies, graviton-graviton interactions are non-existent due to the absence of curvature. The worldline begins to deviate substantially from a rectilinear trajectory at low energies where the uncertainties in its location are large and the associated graviton wavelengths are at macrocosmic scales. In the ground state of spacetime ( $n = \pm 1$) we notice that the metric signature of Eq.(18) becomes negative and that the worldline is rectilinear.

If we compare the quantized metric of Eq.(18) with the Schwarzschild metric we notice that

$$\frac{2}{n^2} = \frac{2GM(r)}{c^2 r}.\tag{19}$$

This yields a relationship between the quantum state of space-time and the amount of baryonic matter embedded within it as follows

$$n^2 = \frac{c^2 r}{GM(r)}.\tag{20}$$

Eq.(20) shows a family of concentric black hole like spherical surfaces of radii $r_n = n^2 GM/c^2$ with corresponding orbital speeds $v_n = c/n$. The innermost stable circular orbit occurs at $n = 1$ or at half the Schwarzschild radius implying that in the Nexus Paradigm the event horizon is half the size predicted in GR. The square term on the left makes it imperative that the mass term on the right remains positive regardless of the positive or negative vibrational modes of space-time explicit in Eq.(10). This resonates well with the positive mass theorems [28-29]. However, the argument presented here is more direct and is a consequence of the quantization of the gravitational field.

Evidently, the Ricci soliton arising from Eq.(20) has an anti-De Sitter topology and to differentiate it from a Ricci soliton of De Sitter topology we label its quantum state as ñ. We can now replace the stress –momentum tensor in Eq.(17) and express the equation as

$$G_{(nk)\mu\nu} = \tilde{n}^2 \Lambda g_{(nk)\mu\nu} + (n^2 - 1)\Lambda g_{(nk)\mu\nu}$$

$$= (\tilde{n}^2 + n^2 - 1)\Lambda g_{(nk)\mu\nu}.\tag{21}$$

Here the complete Einstein's field equations are expressed in purely geometric terms as a compact Einstein manifold. For any quantum state $n$ in which a Ricci soliton has constant curvature, energy is conserved. The right side is a symmetric tensor expressing the quantum/energy state of spacetime. The left side is a form of a laplacian that averages the paths taken by a test particle in a gravitational field of quantum state $n$.

The linearized Eq.(21) is solved by expressing it as another Ricci soliton in the $N$-th quantum state yielding the equation

$$G_{(Nk)\mu\nu} = N^2 \Lambda g_{(Nk)\mu\nu}.\tag{22}$$

The above equation has a solution

$$ds^2 = -\left(1 - \left(\frac{2}{N^2}\right)\right)c^2 dt^2 + \left(1 - \left(\frac{2}{N^2}\right)\right)^{-1} dr^2 + r^2(d\theta^2 + \sin^2\theta d\varphi^2)$$

$$= -\left(1 - \left(\frac{2GM_N}{rc^2}\right)\right)c^2 dt^2 + \left(1 - \left(\frac{2GM_N}{rc^2}\right)\right)^{-1} dr^2 + r^2(d\theta^2 + \sin^2\theta d\varphi^2). \quad (23)$$

Here $M_N(r) = M_B(r) + M_{DM}(r) + M_\Lambda(r)$ where the terms on right represent the baryonic mass, the DM mass and the DE mass enclosed inside a sphere of radius $r$. This yields a metric equation of the form

$$ds^2 = -\left(1 - 2\left(\frac{GM_B}{rc^2} + \frac{H_0 vr}{c^2} - \frac{H_0 cr}{2\pi c^2}\right)\right)c^2 dt^2 + \left(1 - 2\left(\frac{GM_B}{rc^2} + \frac{H_0 vr}{c^2} - \frac{H_0 cr}{2\pi c^2}\right)\right)^{-1} dr^2 + r^2(d\theta^2 + \sin^2\theta d\varphi^2). \quad (24)$$

Where $\frac{GM_{DM}(r)}{r} = v^2 = (H_0 r)^2 = H_0 vr$ and $\frac{GM_\Lambda(r)}{r} = -\frac{H_0 cr}{2\pi}$. The above metric equation leads to the following equation for gravity

$$\frac{d^2 r}{dt^2} = \frac{GM_B}{r^2} + H_0 v - \frac{H_0 c}{2\pi}. \quad (25)$$

The dynamics become non-Newtonian when

$$\frac{GM_B(r)}{r^2} = \frac{H_0}{2\pi} c = \frac{v_n^2}{r}. \quad (26)$$

Under such conditions

$$r = \frac{2\pi v_n^2}{H_0 c}. \quad (27)$$

Substituting for $r$ in Eq.(26) yields

$$v_n^4 = GM_B(r) \frac{H_0}{2\pi} c. \quad (28)$$

This is the Baryonic Tully – Fisher relation. Condition (26) reduces Eq.(25) to

$$\frac{d^2 r}{dt^2} = \frac{dv_n}{dt} = H_0 v_n. \quad (29)$$

From which we obtain the following equations of galactic and cosmic evolution

$$r_n = \frac{1}{H_0} e^{(H_0 t)} (GM_B(r) \frac{H_0}{2\pi} c)^{\frac{1}{4}} \quad = \frac{v_n}{H_0}. \quad (30)$$

$$v_n = e^{(H_0 t)} (GM_B(r) \frac{H_0}{2\pi} c)^{\frac{1}{4}} \quad = H_0 r_n. \quad (31)$$

$$a_n = H_0 e^{(H_0 t)} (GM_B \frac{H_0}{2\pi} c)^{\frac{1}{4}} \quad = H_0 v_n. \quad (32)$$

## 3.0 Quantum Spatio-temporal Dynamics

We proceed to find the complete covariant canonical quantization of Eq.(21). To this end we express the unit displacement vectors $\boldsymbol{e}$ in terms of the Bloch energy functions $\varphi$. The unit vectors are Dirac 4 vector functions.

$$\boldsymbol{e} = \gamma^\mu \boldsymbol{e}_\mu \varphi . \tag{33}$$

The metric coefficents can then be expressed as

$$g^{\mu\nu} = \gamma^\mu \gamma^\nu \boldsymbol{e}_\mu \cdot \boldsymbol{e}_\nu \varphi\varphi$$

$$= \eta^{\mu\nu} \varphi\varphi I_4$$

$$= \eta^{\mu\nu} \varphi\varphi. \tag{34}$$

The fluctuating Minkowski metric arises from the uncertainity in the locality of the unit vectors. The Ricci flow for the vacuum equations is then expressed as follows

$$\partial_t g_{(nk)\mu\nu} = -\frac{1}{3} cr_{HS} G_{(nk)\mu\nu} = -\frac{1}{3} cr_{HS}(n^2 - 1)\Lambda g_{(nk)\mu\nu}$$

$$= -\frac{1}{3} cr_{HS}(n-1)(n+1)\Lambda g_{(nk)\mu\nu} . \tag{35}$$

The term on the right suggests a covariant and contravariant derivative operating on the metric coefficient such that the Ricci flow when expressed in terms of the Bloch functions becomes

$$-i\partial_t \gamma_\mu \varphi_{nk} \gamma_\nu \varphi_{nk} = -\frac{cr_{HS}}{3\pi^2} i\gamma_\mu \nabla^{(n-1)\mu} \varphi_{nk} i\gamma_\nu \nabla_{(n+1)\nu} \varphi_{nk}$$

$$= \frac{cr_{HS}}{12\pi^2} \gamma_\mu \nabla^{(n-1)\mu} \varphi_{nk} \gamma_\nu \nabla_{(n+1)\nu} \varphi_{nk} . \tag{36}$$

where

$$\nabla^{(n-1)\mu} = \partial^{n\mu} - ik^{1\mu} \text{ and } \nabla_{(n+1)\nu} = \partial_{(n\nu)} + ik_{1\nu}$$

The derivative operators on the right are entangled, and the coupling coefficient, $\frac{cr_{HS}}{12\pi^2}$ is an areal speed which is the speed of entanglement with a numerical value of approximately 5.2 square parsecs per second.

The Ricci flow in the presence of baryonic matter is expressed as

$$-i\partial_t \gamma_\mu \varphi_{nk} \gamma_\nu \varphi_{nk} = \frac{cr_{HS}}{12\pi^2} \gamma_\mu \nabla^{(n-1)\mu} \varphi_{nk} \gamma_\nu \nabla_{(n+1)\nu} \varphi_{(nk)} - \tilde{n}^2 H_0 \gamma_\mu \varphi_{nk} \gamma_\nu \varphi_{(nk)} . \tag{37}$$

where $\tilde{n}^2 = \frac{c^2 r_n}{GM} = \frac{c^2}{v^2}$ and $\frac{1}{3} cr_{HS} \Lambda = H_0$

Thus baryonic matter behaves as a heat sink and the vacuum state of space time as a heat source. Gravitational attraction therefore occurs as a flow of space-time in much the same way as heat flows from a heat source to a heat sink. A test particle of baryonic matter flows along with the space-time to the gravitating mass.

Multiplying both sides of Eq.(37) by the reduced Planck constant $\hbar$, while expressing $\varphi\varphi = g$ and factoring out the Minkowski metric yields

$$\hbar i \partial_t g = -\frac{\hbar^2}{m_G}(\partial^{n\mu} - ik^{1\mu})(\partial_{(nv)} + ik_{1v})g + \tilde{n}^2 \hbar H_0 g. \tag{39}$$

here
$$\frac{cr_{HS}}{3} = \frac{cr_{HS}}{3}\frac{\hbar}{\hbar} = \frac{c^2\hbar}{3H_0\hbar} = \frac{\hbar}{m_G}. \tag{40}$$

In the reference frame of the flowing space-time the covariant and contravariant derivatives are null yielding

$$\Box g + k^{1\mu}k_{1\mu}g = \hbar^2 \Box g + p^{1\mu}p_{1\mu}g = \hbar^2 \Box g + m_G^2 c^2 g = 0. \tag{41}$$

Eq.(41) is a 4D Helmholtz equation in which the source of gravitational waves is the $n = \pm 1$ quantum state or ground state of space-time. The equation also implies the existence of a minimum energy/frequency $\frac{m_G c^2}{\hbar}$ in nature. Gravitational waves therefore are carriers of the ground state 4-momentum $p^{1\mu}p_{1\mu}$, a result that contradicts that of Cooperstock [30]. Their frequency is $f = \left(\left(\frac{c^2}{\lambda^2} + H_0^2\right)\right)^{1/2}$. Since Eq.(41) is an expression for Einstein's vacuum equations $G_{\mu\nu} + \Lambda g_{\mu\nu} = 0$ for extremely weak metric perturbations, it therefore localizes the source of DE to the ground state 4-momentum. A similar conclusion was also obtained in Ref.[24].

The 4D Helmholtz equation on a discretized tangential manifold in which the tangential space is discretized into units of $k^{1\mu}$ becomes a random walk equation. This aspect is of importance in describing entropy on the black hole like surfaces of AdS Ricci solitons of radii $r_{\tilde{n}} = \tilde{n}^2 r_g$. Here $r_g$ is the gravitational radius. In [31], it is demonstrated that the black hole like surfaces are marginally stable or zero energy orbitals and that their expectation values are computed as $\langle r_n \rangle = \frac{r_g}{2}[3n^2 - l(l+1)]$. These expectation values are found to be the stable circular orbitals. For black holes, the innermost marginally stable orbit occurs at $n = 2$ $l = 0$ since the $n = 1$ orbital is the actual black hole surface. That is, at radius $r_2 = 4r_g = 2r_s$. Here $r_s$ is the Schwarzschild radius. The expectation value which is the innermost stable circular orbital is therefore $\langle r_n \rangle = 6r_g = 3r_s$. These results are the same as those computed using geometric means in GR.

Along the geodesic, the total energy of the gravitational field or the Hamiltonian of Eq.(39) is reduced to

$$i\hbar \partial_t g = \tilde{n}^2 \hbar H_0 g. \tag{42}$$

The above equation describes quantum harmonic oscillations of the metric with positive energy levels $E_n = ñ^2 \hbar H_0$ which from Eq.(10) we find $E_{max} = 10^{120} E_{min}$. Thus the gravitational field can be described as a system of nested harmonic oscillators in the form of Ricci solitons. More importantly, Eq.(42) helps to define the equation of continuity for the Klein-Gordon equation expressed in Eq.(41) as follows:

$$\frac{1}{c^2}\frac{\partial}{\partial t}\left(g\frac{\partial g^*}{\partial t} - g^*\frac{\partial g}{\partial t}\right) - \nabla(g\nabla g^* - g^*\nabla g) = -\frac{ñ^2 H_0}{c^2}\frac{i\partial}{\partial t}(gg^* - g^*g) - \nabla(g\nabla g^* - g^*\nabla g)$$

$$= \frac{\partial}{\partial t}(gg^* - g^*g) - \frac{3i\hbar}{ñ^2 m_G}\nabla(g\nabla g^* - g^*\nabla g) = 0. \quad (43)$$

The above continuity equation expresses the conservation of information along a geodesic. The information contained in the metric coefficients describes all the possible forms of the geodesic in the quantum state ñ which is determined by the gravitating mass $M$. At high energy states the divergence from a rectilinear trajectory is low and increases with decreasing quantum state. Likewise, the information density increases with increase in quantum state. In other words a test particle is highly localized at high quantum states of space-time and becomes delocalized at low quantum states.

The complete Ricci Flow equation includes a reaction term $g^2$ such that Eq.(42) can be expressed as

$$3i\hbar \partial_t g = 3ñ^2 \hbar H_0 g + im_G c^2 g^2. \quad (44)$$

The Lagrangian of the harmonic system then takes the form
$$\mathcal{L}(g) = -ñ^2 \hbar H_0 \left(g^2 + \left(\frac{m_G c^2}{3ñ^2 \hbar H_0}\right)^2 g^4\right). \quad (45)$$

The potential assumes a Mexican hat morphology under the Wick rotation $g \to ig$

$$\mathcal{L}(g) = -ñ^2 \hbar H_0 \left(-g^2 + \left(\frac{m_G c^2}{3ñ^2 \hbar H_0}\right)^2 g^4\right). \quad (46)$$

Assuming a spring constant $k = 2ñ^2 \hbar H_0$ the Lagrangian then expresses the difficulty to generate excitations of the field at large ñ, but can however be readily generated at ñ = 1. Gravitational wave energy therefore depends mostly on the amplitude of the wave and not on the frequency. The harmonic vibrational modes are a square number series of the fundamental frequency and can be represented as a sequence of binary bits. The $n$-th harmonic is the sum of the $n - 1$ harmonics.

DE is the ground state of the gravitational field and therefore from Eq.(46) must behave like the Higgs field but with negative potential energy minima everywhere. Given that $m_G c^2 = 3\hbar H_0$ therefore when $n = 1$ the vacuum expectation value for $g$ is $\pm\sqrt{\frac{1}{2}}$ implying that the measured value of the cosmological constant $\Lambda' = \Lambda\langle g\rangle \sim 0.707107\Lambda$ where $\Lambda = 3\left(\frac{H_0^2}{c^2}\right)$ is the cosmological constant in a De Sitter vacuum. This yields a theoretical value of $\Lambda' = 1.140903 \times 10^{-52} m^{-2}$ using the Planck 2013 [32] value of the Hubble constant of $H_0 = 2.1927664 \pm 0.0136 \times 10^{-18} s^{-1}$. The theoretical result agrees very well with the measured value by the Planck Collaboration [33] for $\Omega_\Lambda \sim 0.7$. It is worth mentioning that at high energies the second term on the right side of Eq.(46) becomes negligible and the vacuum expectation value at these energies is zero. Thus only the ground state vacuum expectation value contributes to the vacuum energy density. This aspect resolves the cosmological constant problem.

The values of $g$ can only assume positive values because of the constraint $g = \varphi\varphi$. This scenario favors a slow roll of $g$ down the potential hill and depicts a half Mexican hat potential.

The Hamiltonian of Eq.(44) breaks discrete time translation symmetry and generates quantum state reduction in a gravitational field as discussed by Wenzel in Ref.[34]. Here the reaction term is increasing the density of information by driving the system towards low quantum states and therefore confining information to fewer quantum states. The reduction in quantum states applies to quantum matter resulting in baryonic matter transforming into a Bose-Einstein condensate (BEC) at low gravitational quantum states. This result suggests that neutron star matter could be a BEC. The ordered BEC state is the final result of discrete time translation symmetry breaking. Quantum state reduction in low gravitational states could shed some light on the black hole information paradox. Also, recall that the Higgs mechanism was engineered and not derived from first principles but here we show how this mechanism could arise from first principles.

 Concluding remarks and future directions.

Energy-momentum localization in classical GR is a difficult problem, primarily because GR does not include an important aspect of space-time which is the quantum vacuum. Thus only a quantum theory of gravity that provides a link between space-time, gravity and the quantum vacuum can provide a more acceptable description of energy in a gravitational field. In this work, we have translated the geometric language of GR into the wave language of QFT following a slightly different procedure as in Ref.[25]. This quantization procedure enables a complete description of the gravitational field in which DE and DM are a natural aspect of the field equations. The covariant derivatives of the quantum theory describe the affine connection field in which the ground state graviton is the messenger particle with the smallest possible mass-energy in nature. **Gravity appears to be a massive Higgs like scalar field that couples to the Minkowski metric causing it to warp or bend its rectilinear world lines. The coupled state is the graviton.** In synthesis, energy-momentum localization in quantum gravity shows that a free falling object will experience larger vacuum fluctuations (uncertainties in location) in strong gravity than in weak gravity and that the amplitudes of these oscillations define the energy of the free falling object.

Future studies will be centered on studying random walk phenomena and entropy on the tangential manifold using tools such as the Quantum Monte Carlo Method.

**Acknowledgements**

The author acknowledges valuable critique and analysis of the research given by Moletlanyi Tshipa and Christian Corda as well as that from the anonymous reviewers.

**Funding Statement**

We gratefully acknowledge funding and support from the Physics Department of the University of Botswana.

**Data Availability**

Empirical data used in this research can be found in the cited articles.

**Conflict of interest**

We declare no conflict of interest.